\documentstyle[11pt,epsf]{article}
\input psfig
\def\beq{\begin{eqnarray}}
\def\eeq{\end{eqnarray}}
\def\bea{\begin{eqnarray*}}
\def\eea{\end{eqnarray*}}

\def\centeron#1#2{{\setbox0=\hbox{#1}\setbox1=\hbox{#2}\ifdim
\wd1>\wd0\kern.5\wd1\kern-.5\wd0\fi 
\copy0\kern-.5\wd0\kern-.5\wd1\copy1\ifdim\wd0>\wd1 
\kern.5\wd0\kern-.5\wd1\fi}}
\def\ltap{\;\centeron{\raise.35ex\hbox{$<$}}{\lower.65ex\hbox{$\sim$}}\;}
\def\gtap{\;\centeron{\raise.35ex\hbox{$>$}}{\lower.65ex\hbox{$\sim$}}\;}

\def\lsim{\mathrel{\ltap}}

\def\singleandthirdspaced{\baselineskip=\normalbaselineskip\multiply
    \baselineskip by 130\divide\baselineskip by 100}
\def\singlespaced{\baselineskip=\normalbaselineskip}

\newcommand{\newc}{\newcommand}
\newc{\qbar}{{\overline q}}
\newc{\Kahler}{K\"ahler }
\newc{\deltaGS}{\delta_{\rm GS}}
\begin{document}
\begin{titlepage}
\begin{flushright}
{\large hep-th/0407043 \\ SCIPP-2004/11 \\ SU-ITP-04-28 }
\end{flushright}

\vskip 1.2cm

\begin{center}

{\LARGE\bf Low Energy Supersymmetry From the Landscape}

\vskip 1.4cm

{\large  Michael Dine ~and~ Elie Gorbatov} \\ \vskip 0.25cm {\it 
Santa Cruz Institute for Particle Physics,
    Santa Cruz CA 95064  } \\
\vskip 0.6cm {\large Scott Thomas} \\ \vskip 0.25cm {\it Physics 
Department, Stanford University, Stanford, CA 94305 }

\vskip 4pt

\vskip 1.5cm

\begin{abstract}


There has been some debate as to whether the landscape does or 
does not predict low energy supersymmetry.  We argue that under rather mild
assumptions, the landscape seems to favor such breaking, quite possibly
at a very low scale.
Some of the issues which must be addressed in order to settle 
these questions are the relative frequency
with which tree level and non-perturbative effects
generate expectation values for auxiliary fields and the superpotential,
as well as 
the likelihood of both $R$- and non-$R$ discrete or accidental 
symmetries. 
Alternative scenarios with warped compactifications or large extra 
dimensions are also discussed.



\end{abstract}

\end{center}

\vskip 1.0 cm

\end{titlepage}
\setcounter{footnote}{0} \setcounter{page}{2} 
\setcounter{section}{0} \setcounter{subsection}{0} 
\setcounter{subsubsection}{0}

\singleandthirdspaced

\section{Introduction:  Supersymmetry and the Landscape}

Supersymmetry and/or technicolor have long been advocated as 
possible solutions to the hierarchy problem.  But our failure to 
understand the cosmological constant, from the beginning, raised 
the nagging worry that there might be other explanations of 
apparent fine tunings. String theory made low energy 
supersymmetry seem more plausible; string/M theory is replete 
with solutions with $N=1$ supersymmetry to all orders in various 
expansion parameters.  But for many years, the failure to find 
isolated vacua (and again to understand the cosmological 
constant) left considerable uncertainty as to whether low energy 
supersymmetry might be an outcome of string theory.

For better or worse, we have, at present, only one concrete 
proposal to understand the apparent small value of the 
cosmological constant, $\Lambda$, what Weinberg has dubbed the 
``weak anthropic principle" \cite{Weinberg:1988cp} (more recently, 
Weinberg's specific implementation has been called the 
``structure principle" in \cite{Arkani-Hamed:2004fb}).  This is 
the idea that there is a vast ``landscape" of possible stable or 
metastable ground states of the system, in which the cosmological 
constant takes on a ``discretuum" of values.  
Weinberg (and subsequently \cite{Garriga:1999bf}) 
argued that while different regions/periods of the universe might 
exist in different states, only those with sufficiently small 
$\Lambda$ would lead to cosmological formation of complicated structure.

Various suggestions as to how a landscape might arise have been 
made through the years. The most detailed and plausible 
implementation of this program to date has been provided by KKLT 
\cite{Kachru:2003aw}. These authors studied compactifications of 
string theory on Calabi-Yau spaces with fluxes. They noted that 
the presence of fluxes typically fixes the complex structure 
moduli, and argued that generically various effects would fix the 
Kahler moduli as well (further studies of the stabilization of 
Kahler moduli appear in \cite{Denef:2004dm}). In the simplest 
version, the states are supersymmetric, with AdS supersymmetry. 
However, they noted that states with broken supersymmetry are 
likely to exist, at the same level of analysis.

There has been much handwringing about the question of whether 
such a picture can in any sense be viewed as science.  It is not 
clear, for example, that a priori predictions (and falsification) 
are possible.  Douglas has advocated a statistical analysis of 
string vacuum states, and has initiated such an effort with 
various collaborators \cite{Douglas:2003um,Douglas:2004kp} as has 
Kachru and collaborators \cite{Giryavets:2003vd,Giryavets:2004zr}. 
In \cite{Banks:2003es}, some of the present authors considered a 
number of issues which the landscape must confront. Some 
cautionary notes about the very existence of the flux discretuum were 
advanced.  But accepting the picture suggested by effective field 
theory at face value, it was argued that the flux discretuum 
might well be falsifiable, and at the same time, might be 
predictive.  It was stressed that within the flux discretuum, all 
of the parameters of low energy physics -- the particle content, 
gauge groups, coupling constants, etc., vary as one moves around 
in the landscape.  As a result, the question which Weinberg 
raised is enlarged and sharpened.  We might imagine, for example, 
that the hierarchy has the value it has because only if the weak 
scale is close to its current value, can the sort of complex 
astrophysics and 
chemistry required for life arise.  This sort of weak anthropic 
reasoning might also explain the values of other constants, such 
as the gauge couplings.  But it was also noted that there are many 
parameters of low energy physics which do not have any obvious 
implication for the existence of observers.  These should either 
be random variables, typical of the distribution, or they should 
be determined by some underlying principles. While the sharpest 
potential conflict is provided by the small value of $\theta_{\rm 
QCD}$, the masses and mixings of the heavier quarks, the weak gauge 
group, and some cosmological parameters are also puzzling.
Possible problems with symmetry explanations and light moduli 
were discussed.

But if one closes ones eyes to this set of worries, there is an 
interesting possible prediction:  low energy supersymmetry
with visible sector superpartner masses at or near the weak scale. 
Given that SUSY effective field theories are inherent to the KKLT 
construction, one might imagine that this is the easiest -- and 
most urgent -- question to address.  In the spirit of the program 
which has been pursued by Douglas et al and Kachru et al, one 
must ask whether the vast majority of states with, say, small 
cosmological constant and a hierarchy between the weak and 
Planck scales might have supersymmetry broken at comparatively low 
energy.   A tentative argument -- scenario might be a better word 
-- along these lines was put forward in \cite{Banks:2003es}.  
This scenario has been criticized in 
\cite{Susskind:2004uv,Denef:2004ze,Douglas:2004qg}.
In this note, we sharpen the argument that low energy SUSY is a 
likely outcome of the landscape.  Reviewing studies of the landscape
done to date, we outline several scenarios for
more detailed phenomenologies.  These include a rather standard 
hidden sector supergravity scenario and a special case of the proposal of 
of \cite{Arkani-Hamed:2004fb} (which is in fact quite close to a 
proposal put forward long ago in \cite{Dine:1992yw}).  But we
argue that with some rather mild assumptions about discrete symmetries
in the landscape, supersymmetry breaking at the lowest possible
scales is favored (i.e. something like gauge mediation).

In the next section, we briefly review the KKLT analysis, 
focusing on the nature of the approximation, and on 
distinguishing supersymmetric and non-supersymmetric states.  In 
particular, we will explain the sense in which supersymmetry, if 
unbroken at lowest order in a certain small parameter, remains 
unbroken to all orders.  Supersymmetry breaking, then, must be a 
feature of the non-perturbative dynamics of the light fields.  
This implies, we will argue, a distribution of supersymmetry 
breaking scales which is roughly flat on a logarithmic scale. In 
the following section, we will discuss selection issues.  We 
argue, quite generally, that quantities which are not subject to 
anthropic constraints must be typical of the distributions. We 
review why quantities such as $\theta_{\rm QCD}$ do not seem 
typical of the flux discretuum. But laying this concern aside, we 
put forward a pragmatic approach to the anthropic principle. We 
will argue that there are a number of quantities which could well 
be anthropically constrained -- the cosmological constant, the 
dark energy density, the baryon to photon ratio, the gauge 
couplings, the light quark masses, and others.  It is probably 
beyond our present capabilities in science to claim that 
observers exist only at one (or a few) points in this parameter 
space.  Instead, as we explain in section
three, we will take the point of view that nature has 
done the analysis correctly.  We will simply require that the 
cosmological constant, the weak scale, the dark energy density 
and so on lie near their observed values (we refer to this
as the ``Pragmatic Anthropic Principle").  The question of 
predicting supersymmetry is then the question:  are the vast 
majority of states which satisfy these constraints approximately 
supersymmetric (in the sense of low energy supersymmetry).

In the fourth section, we will impose the
constraints on the cosmological constant and weak scales. 
We will see that in the KKLT version of the 
landscape, small cosmological constant and dynamical 
supersymmetry breaking leads to a distribution of supersymmetry 
breaking scales that is roughly uniform on a logarithmic scale.  
Low energy supersymmetry is not particularly favored, but neither 
is it disfavored.  However, requiring also that the Higgs mass be 
near its observed value strongly disfavors supersymmetry breaking 
at scales larger than the intermediate scale, provided the $\mu$ 
term vanishes in the leading approximation.  We argue that the 
cost of a vanishing $\mu$ term might not be severe.  But we also
conjecture that there is likely to be a large set of states in
which, in the leading approximation, supersymmetry is unbroken
with $W=0$, as a result of discrete symmetries.  Dynamical
supersymmetry breaking in some of these states leads to a quite
different distribution of cosmological constants.  If such symmetries
are not too rare on the landscape, 
supersymmetry breaking at the lowest possible energies
is overwhelmingly favored.

Section five is devoted to enumerating pathways to phenomenology
based on these observations.  Very low energy supersymmetry
breaking is suggestive of gauge mediation.  But the higher scale
breaking which seems to result from
the KKLT vacua also suggests interesting possibilities.
We note that the KKLT 
version of the landscape has a cosmological moduli problem, if 
the supersymmetry breaking scale is low.  The Kahler moduli are 
fixed in the KKLT story, but in states with small cosmological 
constant and small supersymmetry breaking scale, their masses are 
comparable to $m_{3/2}$.  Obtaining an acceptable dark matter 
spectrum then argues against a very low scale for supersymmetry 
breaking (less than the intermediate scale). In this version
of the landscape, we 
suggest, the most plausible explanation for the dark matter 
density is that the moduli are heavy, with masses of order $100$ 
TeV.  We review an old argument why in a large class of models, 
dynamical supersymmetry breaking at intermediate scales leads to 
an anomaly mediated spectrum for gauginos (but not squarks and 
sleptons) \cite{Dine:1992yw}. This is, in fact, one of the 
scenarios advocated recently by Arkani-Hamed and Dimopoulos 
\cite{Arkani-Hamed:2004fb}.  Another class of models leads to a 
more conventional picture (but with the usual problems of flavor 
changing processes).

Up to this point, we have contrasted supersymmetric states with 
non-supersymmetric states without any special features (what we 
might call ``randomly broken supersymmetry"(RBSs)). References 
\cite{Denef:2004ze,Giryavets:2004zr} suggest that many of the flux 
vacua lie near conifold points.  This raises the prospect that 
hierarchies may arise through warping.  
Reasons to think that warping or low energy supersymmetry might predominate are 
enumerated. We note, in particular, that stationary points with low 
energy supersymmetry are necessarily local minima of the moduli 
potential, while this is not the case for non-supersymmetric 
conifolds.

In the conclusions, we try to enumerate a set of questions which, 
if answered, would settle the question of whether low energy 
supersymmetry breaking, warping, or RBSS might emerge from the 
landscape.  We explain that many of these questions are accessible.
If low energy supersymmetry is favored, it may even be possible
to determine which if any of the pathways we have listed is favored.
We stress that not only may the landscape be predictive, it may be
falsifiable.  We also stress that within the landscape, conventional
notions of naturalness are sharpened, not abrogated.


\section{Supersymmetric and Non-Supersymmetric Vacua in the Flux Discretuum}

We will take the KKLT analysis as our model for determining 
parameters in the discretuum.  Clearly we do not understand the 
theory well enough to say that the KKLT states exist, much less 
that they are typical.  But this gives us a concrete framework in 
which to work, and to explore possible scenarios.

KKLT imagine self-consistently fixing the values of the moduli. 
They first integrate out Kaluza-Klein modes, and study an 
effective action for complex structure and Kahler moduli.  They 
then show that the superpotential can fix all of the complex 
structure moduli.  Integrating out these fields, leaves an 
effective action for the Kahler moduli (which we denote 
generically by $R$).  This analysis is reliable provided that the 
compactification radii are large. A crucial parameter in this 
action is $W_o$, the expectation value of the superpotential. 
They argue that non-perturbative effects in this action in $R$ 
(i.e. exponentially small for large $R$) will fix the remaining 
Kahler moduli, $R \sim -\ln(W_o)$. So the approximation may be 
reliable precisely in the limit of small $W_o$, which we will see 
is the regime in which low energy supersymmetry breaking may be 
important.

In the simplest version of this story, supersymmetry is unbroken 
at this stage. The small parameter is $W_o$; it determines the 
radii and the masses of the Kahler moduli. Large values of fluxes 
can also serve as small parameters.  In the discretuum, $W_o$ can 
be argued to be distributed roughly uniformly as a complex 
variable \cite{Kachru:2003aw,Denef:2004ze}. Corrections to this 
approximation will not generate SUSY-breaking $F$ or $D$ terms 
for the moduli.  This follows from an argument given long ago by 
Witten \cite{Witten:nf} (suitably generalized to AdS spaces):  in 
order that supersymmetry be broken spontaneously, it is necessary 
that there exist a massless particle which can serve as the 
Goldstino (longitudinal mode of the gravitino). {\it If 
supersymmetry is broken spontaneously, this must be due to the 
dynamics of other light fields, as in field theories which 
exhibit dynamical supersymmetry breaking.}

KKLT went on to consider some possible mechanisms for 
supersymmetry breaking, in particular by considering 
configurations with anti D3 branes.  But for our purposes, what 
is important here is that there is no reason to think there are 
many more non-supersymmetric than supersymmetric vacua at this 
level of approximation.

Within theories in which supersymmetry is broken at lowest order, 
Douglas has argued that supersymmetry breaking is likely to occur 
at the shortest distance scales \cite{Douglas:2004qg}.
A similar argument has been 
made by Susskind 
\cite{Susskind:2004uv}.
Indeed, for such states, it seems likely that the overwhelming number of 
states have supersymmetry broken at the Planck scale, so there 
are simply far more states with small cosmological constant and 
large SUSY breaking than with small SUSY breaking.    

But as long as the number of supersymmetric states is at least 
comparable to the number of non-supersymmetric states, we would 
argue that these are by far the most important. Our experience 
with supersymmetric field theories would suggest that dynamical 
supersymmetry breaking is common.  Indeed, if the landscape has 
anything to do with nature, chiral gauge theories must occur 
frequently, and such theories often break supersymmetry.  
Supersymmetry breaking scales in these states are likely to be 
distributed uniformly on a log scale. We will argue shortly that 
this dynamical breaking is likely to provide far more states that 
satisfy our anthropic requirements than the (random) 
non-supersymmetric ones with tree-level breaking. 

Of course, there could be other constructions of some type which 
produce overwhelmingly more states with broken supersymmetry. 
Fortunately, these are questions we might realistically hope to 
answer over time. For now, our working assumption will be that 
this is not the case.


\section{A Pragmatic Approach to the Anthropic Principle}

The anthropic principle, whatever ones attitude towards it, is 
difficult to implement in practice.  Even the seemingly simple 
question of the cosmological constant raises troubling issues.  
For example, if one only requires the formation of structure, and 
if one allows the dark matter density or the baryon to photon 
ratio or the amplitude of initial perturbations to vary, there is 
a huge range of allowed values of the cosmological constant 
\cite{Aguirre:2001zx}.  We will see quite explicitly in our 
subsequent discussion that these quantities can all vary wildly 
in the landscape. It is possible that more complicated 
considerations involving the detailed nature of galaxies and 
stars sufficiently constrain the parameter space that all of 
these quantities are fixed. Similar statements apply to various 
microscopic quantities, such as the QCD scale, the light quark 
masses, and so on.  

Our approach to selection will be pragmatic (we will refer to 
this as the Pragmatic Anthropic Principle, PAP).  We will assume 
that nature has implemented the anthropic principle correctly, 
and that all quantities which might plausibly be anthropic are 
fixed by some set of considerations to be close to their present 
values (say within 10\% or so).   Specifically, we will assume 
that the following are fixed:
\begin{enumerate}
\item  The cosmological constant \item  The ratio of the weak 
scale to the Planck scale \item  The QCD scale and the gauge 
couplings \item  The dark matter density \item  The baryon to 
photon ratio \item  The primordial fluctuation spectrum
\end{enumerate}
The question is:  {\it  granted this set of assumptions, is it 
possible to make predictions in the landscape; equivalently, is 
the landscape falsifiable.}

There are, as we have argued elsewhere, a number of quantities 
for which it is difficult to conceive of anthropic arguments.  
One example is the QCD $\theta$ parameter \cite{donoghue}.
 In the KKLT version 
of the landscape, it is necessary that the small value of 
$\theta_{\rm QCD}$ somehow be correlated with some other 
anthropic requirement. Perhaps, for example, it is necessary that 
axions provide the dark matter.  This might explain the presence 
of a light axion.  Perhaps, for example,
some modulus is not fixed as in KKLT, but instead
is only stabilized by low energy dynamics, leaving
a light axion; this might be selected for by
cosmological considerations.
Another possibility, 
is that an $R$-axion common to many models of dynamical supersymmetry 
breaking acts as the QCD axion. By itself, neither of these
ideas is adequate to 
explain the smallness of $\theta_{\rm QCD}$, however.  Additional 
discrete symmetries, or something else, are probably required. It 
is quite possible that the whole landscape picture might be 
falsified on such grounds.  
For now, we will assume that there is a solution of this and similar 
problems.  We will focus, instead, on supersymmetry.  Since 
supersymmetry is an essential feature of the KKLT construction 
(and most string constructions), it would seem that the question 
of low energy supersymmetry would be the most straightforward to 
settle. Douglas and Susskind have argued that there is no 
prediction of low energy supersymmetry in the landscape.\footnote{More
recently, both authors have realized that low energy dynamical
breaking might allow such a prediction.}  If this 
is the case, it is quite disappointing.  
It would suggest that the 
landscape may make no distinctive verifiable predictions, 
and that we have no theoretical or 
experimental access to the questions we have long viewed as 
fundamental in particle physics. But as we have explained, this 
argument is based on looking at a set of states which are not 
likely to be particularly important (but they {\it are}, as 
Douglas has stressed, the states for which the most 
straightforward analysis is currently possible).


\section{Is Supersymmetry a Feature of the Landscape?}

 \subsection{Supersymmetry and Discrete Symmetries}
 
 The landscape, whether it leads to supersymmetry or not, almost certainly
 requires discrete symmetries to be consistent with the requirements
 we have enumerated above.  For example, the authors of \cite{Davoudiasl:2004be} 
 list a minimal extension of the Standard Model to account for 
 the requirements given above.
  This minimal model includes a light, stable scalar which serves
  as the dark matter.  This particle is assumed stable
  due to discrete symmetries (and it is light through tuning).  
  An alternative dark matter
  candidate would be an axion, but if the axion is to be sufficiently light without
  extraordinary fine tuning (and, again, it is not clear what 
  anthropic consideration
  would require this), one needs elaborate discrete symmetries.  
  Other features of the
  low energy theory probably require symmetries as well.
  
  In the case of supersymmetry, the need for discrete symmetries is at least
  as urgent.
  As noted in \cite{Banks:2003es}, anthropic considerations 
  do not account for the long life of the
  proton, so something analogous to $R$-parity is probably required.  
  Discrete symmetries
  might also be required to understand the smallness of the $\mu$ term (it would be
  interesting to know whether, as in weakly coupled strings, light fields arise in
  the flux discretuum without such symmetries).  As we will discuss later,
  in warped scenarios, the need for discrete symmetries may be more urgent.
  
  Discrete symmetries are ubiquitous in weakly coupled strings.  Typical
  toroidal or Calabi-Yau compactifications have large discrete symmetry
  groups at points in the moduli space.    It is clear that they
  can also arise in the landscape, but that
  there is some price to be paid; in the flux vacua, some
  number of fluxes must vanish.  In \cite{Banks:2003es}, some 
  simplified models were
  examined in order to see what fraction of states might be expected to possess
  useful discrete symmetries.  The results were not conclusive, 
  and the models not sufficiently
  realistic to make a reliable assessment.  Kachru and collaborators are 
  currently examining
  this question in more realistic models, and tentatively conclude 
  that the price is not
  high (i.e. not a huge exponential factor) \cite{kachruprivate}.
    

\subsection{Is Supersymmetry Favored}

 In \cite{Banks:2003es}, a scenario was outlined in which 
 low energy supersymmetry might emerge
 as a prediction of the landscape.  Here we will consider 
 this type of argument in more
 detail.  As stressed above, our focus will be on dynamical 
 supersymmetry breaking
 in the supersymmetric states of KKLT.  This was, in fact, the focus 
 of \cite{Banks:2003es}.
 We will see that the argument given there that 
a uniform distribution of $W_o$ along with the constraint 
of a small cosmological
 constant favors low energy supersymmetry is not correct.  
 Instead, with these assumptions one predicts that
 the supersymmetry breaking scale has a flat distribution 
 on a logarithmic scale.

We model dynamical supersymmetry breaking by supposing that the 
scale of supersymmetry breaking is $\mu^2 = e^{-{8\pi^2 / g^2}}$. 
In addition, we suppose that the distribution of $g^2$ 
is reasonably flat \cite{Denef:2004ze}.\footnote{We 
thank M. Douglas and L. Susskind for discussions of this
point.}  
This is in fact the case for   
vacua near a conifold for which the distance from the conifold point
is dual to a scale generated dynamically by dimensional 
transmutation \cite{Denef:2004ze,Giryavets:2004zr}.
We want to ask the probability that 
$\Lambda < \Lambda_o$ (to simplify, we will take the cosmological 
constant to be positive).  We suppose that \beq \Lambda = \mu^4 - 
3 \vert W_o \vert^2
\eeq
where $M_p=1$, and for the observed value 
of the vacuum energy $m_{3/2} \sim |W_o|$. Calling $F_1(\Lambda_o)$ 
the fraction of states with $\Lambda <\Lambda_o$, $P_W(W_o)$ the 
probability density for $W_o$, and 
$P_{g^2}(g^2)$ 
the probability density for the appropriate 
coupling, we have:
\beq F_1(\Lambda<\Lambda_o) = \int_0^{W_{\rm 
max}} d^2 W_o P_W(W_o) \int_{\ln(\vert W_o\vert^2)}^{\ln(\vert W_o 
\vert^2 + \Lambda_o)}
d(g^2)P_{g^2}(g^2)
\label{FLambda}
\eeq
$$~~~~~\approx \int_0^{W_{\rm max}} d^2 W_o 
{\Lambda_o \over \vert W_o \vert^2} P_W(W_o){1 \over \ln{W_o}^2} 
P_{g^2}(-1/\ln(W_o))
$$
So we see that if $P_W$ is reasonably flat near the origin, as is 
$P_{g^2}$ then this condition, by itself, predicts a 
distribution of supersymmetry breaking which is flat on a log scale. Note 
that the prefactor in the second line of (\ref{FLambda}) for the 
total fraction of vacua with vacuum energy $\Lambda < \Lambda_o$ 
is the vacuum energy $\Lambda_o$ itself. So the tuning associated 
with the observed value of the vacuum energy for this class of 
vacua with uniformaly distributed $W_o$ and dynamically broken 
supersymmetry is then roughly $\Lambda_o \sim 10^{-120}$ per 
decade.

Considering the question of the gauge hierarchy, however, does 
strongly disfavor very large values of $W_o$. 
Now, 
if the vacuum energy and weak scales are uncorrelated, 
we might expect that the fraction of states 
with fixed Higgs mass $m_H$ for $W_o \sim m_{3/2} \gg m_H$, would 
be suppressed by at least $m_H^2 \over W_o^2$ because of the 
requisite tuning of the Higgs potential. 
However, for $m_H^2 M_p\lsim 
W_o \lsim m_H M_p^2$ there is not necessarily any additional tuning 
since for any $W_o$ in this range there is in principle always 
some messenger sector which gives $m_H$ of order the electroweak 
scale. So if we call $F_2$ the fraction of states with
a suitable $\Lambda$ and Higgs
mass:
 $F_2(\Lambda<\Lambda_o; 10~{\rm GeV} < m_H 
<1~{\rm TeV})$, say, this looks like:
\beq
F_2(\Lambda<\Lambda_o; 
10~{\rm GeV} < m_H <1~{\rm TeV}) \sim \left\{
\begin{array}{cl}
\Lambda_o\int^{W_{\rm max}}_{W_{\rm int}} d^2 W_o 
   {{\rm TeV}^2 \over \vert W_o \vert4} & 
    ~~~~~~W_o > W_{\rm int} \nonumber \\
   F_1(\Lambda < \Lambda_o) & 
    ~~~~~~W_o < W_{\rm int} \\
\end{array}
\right. \eeq where 
$W_{\rm int} \sim m_{3/2} M_p^2$ is the 
intermediate scale. So supersymmetry breaking well above the 
intermediate scale is strongly disfavored. Distinquishing a 
prefered supersymmetry breaking scale within the range $m_H^2 M_p
\lsim W_o \lsim m_H M_p^2$ requires answering the harder question of 
what type of messenger sectors are most common on this part  
of the landscape. 


\subsection{A Possibility for a Huge Enhancement}

In flux vacua constructions, 
the distribution of vacua is roughly uniform in 
$W_o$ once the moduli
are fixed.  From the point of view of obtaining
models with a controlled approximation,
a small non-zero $W$ is a virtue, because it leads to large
values of the radii, and showing that some 
fraction of states had this property was one
of the successes of
KKLT.  However, there may well be a substantial subset where $W_o$ vanishes,
due to an unbroken, discrete $R$-symmetry (supersymmetry is still preserved).
In such vacua, the Kahler moduli will generically be of order
one and there are no obvious small
parameters.  But one expects
that dynamical supersymmetry breaking in the 
low energy theory will sometimes
generate both non-vanishing
$F$ (and $D$) terms, as well as a non-vanishing $W$.  If there are a discretuum
of such states, then it is potentially much easier to cancel the cosmological
constant than in the states with a flat distribution of  $W_o$.  
Here it is helpful
to put back the Planck scale, and note that if $\mu$ is the dynamical
scale,
\beq
\vert DW \vert^2 \sim \mu^4; ~3 \vert W \vert ^2 \sim \mu^2 M_p^2.
\eeq
These terms are of the same order if $\mu^2 \sim m_{3/2} M_p$, i.e.
for intermediate scale supersymmetry breaking (this is a point which T. Banks
has stressed for many years).  So, roughly speaking, among states with
intermediate scale dynamical supersymmetry breaking and vanishing $W_o$,
a fraction 
$\Lambda \over M_{int}^4 $ have suitable cosmological constant,
as opposed to ${\Lambda \over M_p^4}$.  This is a huge advantage for these
states, and may well win over the suppression required to obtain such
a discrete symmetry.

But in fact, this sort of reasoning {\it inevitably suggests that supersymmetry
should be broken at the lowest possible scale.}  Suppose, for example, 
that in addition
to the supersymmetry breaking sector, the theory contains another 
strongly interacting
gauge theory which dynamically breaks the $R$-symmetry but not supersymmetry 
(as gluino condensation
does).  This will generate a constant superpotential, with a distribution
essentially flat on a log scale.  So if $\mu$ is the supersymmetry breaking
scale, cancelling the cosmological constant favors low $\mu$ by $\mu^{-4}$,
modulated by logarithms of $\mu$.  This would tend to favor a phenomenology
more like gauge mediation.
For regions of the landscape with both $W$ and $DW$ generated dynamically 
the fraction of such states with small enough cosmological constant 
and appropriate weak scale could be as large as $10^{-60}$ for a supersymmetry 
breaking scale just above the weak scale, modulo any suppressions for 
additional symmetries, etc. 

This would be one strong selection principle in favor of discrete $R$ symmetries,
broken or partially broken only at scales well below the Planck scale.  Apart from
the question of the cosmological constant, this could also be relevant to
proton decay, the $\mu$ problem, and dark matter.  


\section{Paths to a Detailed Phenomenology}

\subsection{Gauge Mediation}

In the previous section, we have seen that supersymmetry breaking
at the lowest possible energies (consistent, presumably, with various
selection criteria and also the existence of an adequate density
of states) is likely to be favored in the landscape.  If this is
actually the case, then something like gauge mediation
seems the most likely outcome.  This is in some sense a positive
and exciting result, but it raises concerns.  It is positive in that
it provides the simplest explanation of how problems of flavor might
be resolved in the landscape context.  But it is troubling because
gauge mediated scenarios at this stage tend to be somewhat tuned.
One question is whether some selection criteria 
such as the ones listed in section 3 might explain some
of this tuning when convoluted with features of the distribution of states 
on the landscape. 
We will not provide any answer to this question here.

In the rest of this section, we outline two other plausible 
scenarios for supersymmetry phenomenology.  The first one has 
gauginos in the TeV energy range, but scalars at 100's of TeV or 
more (similar to one of the proposals in 
\cite{Arkani-Hamed:2004fb}).  The second is a more conventional 
supergravity scenario.    We should stress that it is
possible that the landscape, as it is better
understood, will make predictions which one can reliably
say are in gross
conflict or agreement with experiment.


\subsection{KKLT and the Moduli Problem}

If we take literally the construction outlined by KKLT, then 
coupled with a prediction of low energy supersymmetry, we also
encounter a moduli problem.  In the KKLT construction, the Kahler
moduli (radii, etc.) are of order:
\beq
R \approx -c \ln(W_o)
\eeq
and their masses are of order $W_o$.  So the masses of the 
complex structure moduli, in this class of models are of order $m_{3/2}$.
Such models will then, potentially, have a moduli problem.  If $m_{3/2}$
is very small, as in low energy gauge mediation, such moduli would
be particularly problematic.  The PAP is likely
to heavily disfavor such states\footnote{We certainly
don't know how to treat this problem completely,
but requiring, for example, that the initial values
of the fields be sufficiently small that the
moduli don't dominate the energy before the usual time of matter-radiation
equality
could easily cost a factor of $10^{12}$.}.  If $m_{3/2}$ is of order TeV, one has
a more conventional moduli problem; this moduli problem would be solved
if the moduli masses were $100$ TeV or so.  We will see in the next
subsection how such a picture might plausibly emerge from the landscape,
and lead to an interesting phenomenology.

On the other hand, KKLT's requirement of large $R$ came, in part,
from requiring calculability.  But this is not something that nature
requires, and we might imagine that, for generic superpotentials,
the Kahler moduli will also be fixed at small radii and large mass.
In particular, KKLT took the superpotential to include only a single
exponential in these radii, but this is unlikely to be the complete superpotential.
In the radii are small and masses are large, there would be no moduli problem.
As we will discuss below, this leads to more conventional supergravity
scenarios (with their potential flavor problems).


\subsection{A Higher Energy Scenario}

In this section, we will again take the KKLT construction literally.
As we just noted, these vacua potentially suffer from 
a moduli problem. The Kahler moduli, in particular, have masses 
of order $W_o$. 

In the context of ``gravity mediation", the question 
of obtaining intermediate scale breaking dynamically was first 
addressed in \cite{Affleck:xz}.  These authors noted that one 
could readily couple a theory exhibiting DSB to supergravity, and 
obtain masses for scalars.  At that time, the known models of DSB 
did not contain singlet fields, so it was is difficult to obtain 
gaugino masses. Subsequently, it was realized that anomaly 
mediated contributions to gaugino masses \cite{Dine:1992yw} would 
arise at one loop, but it was argued that such models would be 
finely tuned, given the experimental constraints on the gaugino 
masses and the value of the weak scale.  It was also subsequently 
recognized that it is possible to construct models of dynamical 
supersymmetry breaking with singlet fields \cite{Intriligator:1996pu}.
However, theories without singlets suggest a definite phenomenology
which we briefly describe now; those with singlets
allow an alternative which will be discussed
in the next subsection.

In the landscape, the apparent tuning associated with 
heavy scalars which was previously so 
troubling is not relevant if the density of states is higher for 
a larger mass scale, or if selection criteria strongly favor the 
apparently tuned states.

In the present context, the most relevant selection criteria may 
well to be cosmological.   Consider the questions of dark matter 
and the baryon asymmetry.  As Moroi and Randall have stressed, in 
a situation like this, the LSP is likely to be the wino 
\cite{Moroi:1999zb}.  We have noted that there is probably a 
period of moduli domination;
but neutralino dark matter can readily be produced in the decays 
of moduli \cite{lateinf}. A cosmology consistent with observation 
(but not necessarily forced by anthropic considerations) has the 
moduli and gravitino at $100$ TeV or so in order to obtain 
standard nucleosynethesis, with anomaly mediated gaugino masses 
in the few hundred GeV range. Baryons could be produced by the AD 
mechanism \cite{Dine:1995kz}.  This requires that some moduli in 
fact carry baryon number.  Baryons might also be produced by 
baryon violating processes in the decays of the moduli 
\cite{Banks:1993en}. The moduli decays and the AD process both 
introduce several new parameters into the story.  It is plausible 
that this structure, within the framework of the PAP, might 
actually favor visible sector supersymmetry breaking (i.e. scalar 
and gravitino masses) at the 100 TeV scale.  
But note that scalars this heavy require a tuning of order 
at least roughly $10^{-6}$ in the Higgs potential in order to obtain 
acceptable electroweak symmetry breaking. 
So it is quite 
possible that the landscape favors a lower scale, again raising an 
opportunity to falsify the picture.  For the moment, we 
can at least say that the facts of nucleosynthesis are consistent 
with such a spectrum.

We should note that there are further opportunities to falsify this scenario.
The $100$ TeV scale is not quite high enough scale to 
completely solve all the flavor problems of supersymmetric 
theories (though it does solve the CP conserving problems).  
Perhaps there is some combination of anthropic arguments which 
forces the scale higher, or imposes approximate symmetries which 
adequately suppress various flavor violating processes.  But it 
is also possible that there are not, and that this scenario 
within the landscape can be ruled out.


\subsection{A More Conventional Scenario}

It has been known for some time that one can construct models 
with gauge singlets which dynamically break supersymmetry even at 
the renormalizable level \cite{Intriligator:1996pu}.   The 
couplings of these singlets are restricted.  If this arises from 
discrete symmetries, gaugino masses may be suppressed, and we 
will be back to the picture of the previous subsection. But the 
required structure of couplings might arise because the singlets 
are approximate moduli, or perhaps some couplings are simply 
small, again through selection effects (this might not be 
particularly costly; the couplings don't need to be 
extraordinarily small). In this case, gaugino masses could be of 
the same order as scalar masses, leading to a more conventional 
supergravity phenomenology.  In this case, however, the problems 
of flavor are potentially far more severe, as are the moduli 
problems. So again, there are opportunities to falsify this 
scenario within the landscape.  


\section{Conifolds and Warping}

We have seen that low energy supersymmetry is likely to win over 
random non-supersymmetric configurations in the landscape.  It 
has been noted, however, that generically flux vacua may arise 
near conifold points \cite{Denef:2004ze,Giryavets:2004zr}. This 
may provide an alternative to supersymmetry to obtain hierarchies 
\cite{Giddings:2001yu}.

In fact, it would appear that one pays roughly the same price for 
such a hierarchy near a conifold point as one pays for dynamical 
supersymmetry breaking with a uniform distribution of $W_o \ne 0$:
first, a factor of $10^{-120}$ to explain the smallness of 
$\Lambda$, second, an order one factor to explain the gauge 
hierarchy, and third, some factor to obtain discrete symmetries, in 
order to suppress proton decay, obtain dark matter, etc.

One issue which may favor supersymmetry concerns the question of 
unstable (tachyonic) directions in the moduli mass matrix.  In 
approximately supersymmetric vacua with $W_o \ll 1,\Lambda \ll 
\vert W_o\vert^2$, the moduli generically have large, positive 
mass-squared.  In contrast, in the overwhelming majority of 
non-supersymmetric stationary points of the effective potential 
with large $W_o$ and small cosmological constant, one expects 
tachyons.  
We have seen, however, that if $W_o$ is zero at tree-level, 
then supersymmetry wins by an overwhelming amount.
So it should be possible to settle the question of whether
supersymmetry or non-supersymmetric conifolds are favored.

It is difficult to be quite so specific about the
phenomenological issuses
here as in the supersymmetric case.  Proton decay is potentially 
a serious problem.  If the energy scale of the Standard Model 
arises entirely through warping, it is necessary, most urgently, 
to understand why baryon number violating operators up to very 
high dimension are suppressed. One could imagine that this occurs 
through discrete symmetries, and that vacua with a high degree of 
symmetry are selected by anthropic considerations. However, {\it 
anthropic selection probably does not require that the proton 
lifetime be nearly as large as it is observed to be.}  In other 
words, to be compatible with the current limit on the proton 
lifetime, it is necessary to suppress operators through dimension 
$13$ or so; on the other hand, if the anthropic limit on the 
proton lifetime is $10^{16}$ years, it is only necessary to 
suppress operators through dimension $11$.  One could imagine 
that there would be many more vacua capable of suppressing the 
latter set of operators than the former. So typical, anthropically 
selected vacua {\it might} well be inconsistent with 
observation.  Other proposed explanations of the long proton 
lifetime face similar potential difficulties.

This is in contrast to the situation in supersymmetric vacua, 
with at least an unbroken $R$-parity where 
it is necessary 
to suppress a range of dimension five operators, but already 
dimension six operators are not problematic.

This is clearly an issue which requires further study.  It could 
well be that there are simply many more conifold type theories, 
with suitable properties, than supersymmetric theories, or vice 
versa; it could also be that some cosmological considerations 
favor one set or the other.  
For example, we have advanced a 
possible scenario in which the SUSY breaking scale might be 
larger than a TeV, in order to allow suitable structure formation 
in the universe.  
This argument was not compelling; more knowledge about 
these types of vacua is necessary. 
Similarly, in the 
conifold case, it is not clear, at first sight, why we couldn't 
obtain a suitable dark matter density if the scale was one TeV.  
In this case, flavor changing processes would likely falsify this 
possibility.  However, it might be easier to suppress proton decay 
with a higher scale, or perhaps for some not quite so obvious 
reason it is easier to make a suitable dark matter density with a 
higher scale.  These questions are difficult to resolve without
much more detailed information about the distributions.

Finally, we should note in the original KKLT story, the breaking 
of supersymmetry by anti D3 branes located down the throat is 
perhaps to be thought of as the dual of a picture of 
field-theoretic dynamical supersymmetry breaking, and so fits in 
with the discussion of the previous section.\footnote{We thank L. 
Susskind for discussions of this point}.

It is also possible to consider scenarios with large flat extra dimensions
without warping. 
In such scenarios it might be possible to imagine that the fundamental 
scale is just above the weak scale, leading to a fraction of such 
such states with sufficiently 
small cosmological constant as large as $10^{-60}$.  
However, avoiding an additional price for the small four-dimensional Planck scale
would require natural mechanisms for obtaining large volume (this might
be obtained with two large dimensions).  
And the problem of proton stability in generic vacua of this type 
also must be addressed, just as in the case with warping. 

\section{Conclusions}

We have advanced a scenario, which we view as quite plausible, in 
which the landscape predicts low energy supersymmetry, possibly 
of a quite specific form.  But we should stress that this is only 
a scenario, and we have described ways in which supersymmetry 
might not emerge as a prediction. We have noted that 
supersymmetry may be too costly:  there might be far fewer states 
with the requisite discrete symmetries to prevent proton decay 
then there are non-supersymmetric states.  A second possibility, as we 
discussed in the previous section, is that warped geometries are 
sufficiently more numerous than states with dynamical 
supersymmetry breaking.  Some pitfalls with the conifold scenario 
were noted, 
but it is not yet possible to argue decisively for the relative 
importance of one or another of these classes of vacua on the landscape. 

If the following questions were answered, one 
could establish that supersymmetry is or is not a likely outcome of the 
landscape:
\begin{enumerate}
\item  Are there, in the leading tree-level approximation, exponentially 
more non-supersymmetric than supersymmetric vacua?  We have 
indicated that the answer to this question is likely to be no, 
but we certainly cannot claim to have proven such a statement.  
This would favor low energy supersymmetry. 
\item  What is the 
price of discrete symmetries?  In particular,
we need to compare the cost of suppressing proton
decay and 
(if necessary) obtaining a small $\mu$ term with 
the price of light Higgs without supersymmetry ($10^{-36}$ or 
so), times the price of obtaining a stable, light dark matter 
particle (unknown, but probably not less than $10^{-36}$), times 
the other tunings required to obtain an acceptable cosmology.  
\item  Is there a huge 
price for obtaining theories with low energy dynamical 
supersymmetry breaking?  Given the presumption that one can 
obtain a landscape of models with complicated gauge groups and 
chiral matter, it is hard to imagine that the price is enormous 
(in landscape terms).  A part in a billion, for example, would 
likely lead to a prediction of low energy supersymmetry. 
\item Are unbroken discrete {\it $R$-symmetries} at the high scale 
common?  If so, $\langle W \rangle$ must be generated dynamically at
low energies in 
such vacua. 
In this case, we have seen that SUSY breaking at the lowest 
possible scale may be favored. 
\item  
Within the present knowledge of the landscape, non-supersymmetric
conifolds appear to be the most promising alternative to low energy
supersymmetry.  What is the relative abundance of such states compared
to supersymmetric states?
\end{enumerate}

Related to the question of discrete symmetries, it would be 
interesting to know whether in supersymmetric states in the 
landscape, one often obtains massless, vector-like states (i.e. 
whether one can obtain, as in weakly coupled string theory, a 
small $\mu$ term without symmetries).

Answering these questions would establish whether supersymmetric vacua,
as opposed to 
random non-supersymmetric vacua, are far more likely to satisfy 
the anthropic constraints we have enumerated. If the price of vanishing
$W_o$ is not large, we have seen that supersymmetry breaking
at the lowest possible scale wins.

More detailed information about the statistics of the landscape could further
narrow the possibilities for string phenomenology.
In particular, we have
seen that information about the numbers of supersymmetry breaking
low energy theories of various types could lead to
quite specific predictions.

There are various ways the theory could 
fail to predict supersymmetry.  If, for example, $W_o$ is
non-zero at leading tree-level approximation 
in the overwhelming number of states
and one has to understand the smallness of 
the $\mu$ term as a consequence of tuning, this might eliminate 
any possible prediction of low energy supersymmetry.
If it should turn out, with further exploration of the landscape,
that there are vastly more non-supersymmetric than supersymmetric
states, again low energy supersymmetry is not a likely
outcome of string theory.

We should also stress that even in the landscape, 
ordinary notions of naturalness are likely to be relevant.  For example,
we have seen that in the intermediate scale
scenario for supersymmetry breaking 
primordial nucleosynthesis forces gaugino masses into the TeV 
region, with scalars in the $100$ TeV range.  This is certainly
``tuned"; there are almost certainly far more states
with light Higgs particles with supersymmetry broken
at lower energies.   We have suggested a way 
in which this mini hierarchy might arise from selection criteria of a 
cosmological nature.  We leave for further work
the question of tuning in the very low energy
scenario, but note that in this
case, in addition to cosmological concerns,
the need to account for the $\mu$ and $B_{\mu}$ parameters
might explain requisite tunings.  We have also noted that for the conifold 
scenario, proton decay and flavor
violation require coincidences not readily explained
by anthropic reasoning.

The questions in this list seem accessible to investigation.  In 
the background, however, there lurk other questions:  
are these states real and accessible cosmologically?  If they 
are, does the history of the universe prefer some states over 
others?

\noindent {\bf Acknowledgements:} This work supported by the U.S. 
Department of Energy.  We thank A. Aguirre, Tom Banks, Michael Douglas, Willy Fischler, 
Shamit Kachru, and Lenny Susskind for discussions and protecting 
us from worse blunders than those we may have committed above. \noindent 
The work of M.D. and E.G. was supported in part by the US 
Department of Energy. The work of S.T. was supported in part by 
the US National Science Foundation under grant PHY02-44728.


\end{document}